\documentclass[amsfonts,pra,twocolumn]{revtex4}
\usepackage{graphicx}
\usepackage{color}
\usepackage{bm}

\def\de{\mathrm{d}}
\def\mrm{\mathrm}
\begin{document}
\title{Temperature dependence of binary and ternary recombination of $\mathrm{H}_3^+$ ions with electrons}
\author{J. Glos\'ik}
\author{R. Pla\v sil}
\author{I. Korolov}
\author{T. Kotr\'ik}
\author{O. Novotn\'y}
\author{P. Hlavenka}
\author{P. Dohnal}
\author{J. Varju}
\affiliation{Charles University, Mathematics and Physics Faculty, Prague 8, Czech 
Republic}
\author{V. Kokoouline}
\affiliation{Department of Physics, University of Central Florida, Orlando, Florida 32816, USA}
\author{Chris H. Greene}
\affiliation{Department of Physics and JILA, University of Colorado, Boulder, Colorado 80309-0440, USA}

\date{\today}

\begin{abstract}
We study binary and the recently discovered process of ternary He-assisted
recombination of $\mathrm{H}_3^+$ ions with electrons in a low temperature
afterglow plasma. The experiments are carried out over a broad range of
pressures and temperatures of an afterglow plasma in a helium buffer gas. Binary
and He-assisted ternary recombination are observed and the corresponding
recombination rate coefficients are extracted for temperatures from 77~K to
330~K. We describe the observed ternary recombination as a two-step mechanism: First, a rotationally-excited long-lived neutral molecule $\mathrm{H}_3^*$ is formed in electron-$\mathrm{H}_3^+$
collisions. Second, the $\mathrm{H}_3^*$ molecule collides with a
helium atom that leads to the formation of a very long-lived Rydberg state with high orbital momentum. We present calculations of the lifetimes of $\mathrm{H}_3^*$ and of the ternary
recombination rate coefficients for para and ortho-$\mathrm{H}_3^+$. The
calculations show a large difference between the ternary recombination rate
coefficients of ortho and para-$\mathrm{H}_3^+$ at temperatures below 300~K.
The measured binary and ternary rate coefficients are in reasonable agreement with the calculated values.
\end{abstract}


\maketitle

\section{Introduction}
Electron scattering by the simplest polyatomic ion H$_3^+$ is of fundamental
importance in plasma physics because $\mathrm{H}_3^+$ ions are dominant in
many hydrogen-containing plasmas, including astrophysically-relevant plasmas in particular. For
quantum theory, the electron-H$_3^+$ interaction is important because the process involves the simplest 
polyatomic ion that can be studied from first principles without adjustable parameters. In the long history of research on recombination of
$\mathrm{H}_3^+$ with an electron there have been numerous exciting, contradictory and sometimes unaccepted results. We mention here only a few recent reviews devoted to H$_3^+$ recombination at low temperatures (see e.g. \cite{oka2006,geballe2006akm,plasil2002ais,johnsen2005,larsson2008}). A successful theory of $\mathrm{H}_3^+$ recombination with electrons at scattering energies below 1 eV was
developed relatively recently when the Jahn-Teller non-Born-Oppenheimer coupling was included into theory \cite{kokoouline2001mdh,kokoouline2003utt,dossantos2007drh,jungen09}. The measurements of the recombination rate constant in different storage ring experiments have also converged to the same value recently, after it became understood that the internal rotational and vibrational degrees of freedom of the H$_3^+$ ion influence the recombination rate \cite{dossantos2007drh,kreckel2005hrd,mccall2003mn}. The theoretical developments and improvements in storage ring experiments have resulted in a reconciliation of theory and experiment for the binary electron-H$_3^+$ recombination process.

However, the $\mathrm{H}_3^+$ and $\mathrm{D}_3^+$ recombination experiments carried out in afterglow plasmas \cite{plasil2002ais,johnsen2005,larsson2008,glosik2000rhi,glosik2001esr,amano1990drr,Gougousi1995131,smith1993drh,poterya2002rd}  have repeatedly given rate coefficients very different from the ones obtained in the aforementioned storage ring experiments and the theoretical calculations. Until now, the plasma studies have not been fully understood and, as a result, they have been frequently rejected because they do
not mesh with the present understanding of the binary DR process (see, e.g., the very recent review discussing this subject \cite{larsson2008drh}). However, the experimental plasma results are reproducible and they demand an understanding and an integration into the full picture of H$_3^+$ recombination with electrons.  The present work discusses and, we hope, clarifies some aspects of this complex problem.

The plasma measurements are usually carried out in a He/Ar/$\mathrm{H}_2$ gas mixture
(see the review by Plasil \textit{et al.} \cite{plasil2002ais}) or in a pure $\mathrm{H}_2$ gas \cite{amano1990drr,Feher1994}. The main question is how to ``reconcile'' \cite{smith1993drh} the rate constant (including its dependence on experimental conditions) observed in an $\mathrm{H}_3^+$ dominated plasma with actual theory and with data from the storage ring experiments \cite{larsson2008drh}.  The plasma experiments are typically carried out with helium buffer gas at pressures in the range 50--2000 Pa. It has been generally accepted that such pressures are too small to produce appreciable ternary helium-assisted recombination of H$_3^+$. 
The fact that the neutral-stabilized recombination can sometimes play a role was predicted many years ago by Bates
and Khare \cite{bates1965rpi} and confirmed for some ions (but not for $\mathrm{H}_3^+$) by experiments
\cite{cao1991nse,gousset1977ec}. The typical value for the three-body recombination rate coefficient  $K_\mathrm{He}$ with helium as an ambient gas obtained in experiment and estimated theoretically \cite{cao1991nse,gousset1977ec} is of order of $10^{-27}\ \mathrm{cm}^6\mathrm{s}^{-1}$ at 300~K.  Thus, at
pressures of 1300~Pa, the corresponding apparent binary recombination rate
coefficient is smaller than $10^{-9}\mathrm{cm}^3\mathrm{s}^{-1}$, which would therefore be negligible in comparison with
the now-accepted binary H$_3^+ $ recombination rate coefficient  \cite{dossantos2007drh,kreckel2005hrd,mccall2003mn,larsson2008drh,jungen09}.  This would also be negligible in comparison with the binary rate coefficients (about $ 10^{-7}\ \mathrm{cm}^3\mathrm{s}^{-1}$ at 300~K) for the majority of molecular ions. This is the reason why the ternary
recombination of $\mathrm{H}_3^+$ had previously been neglected in FALP (flowing afterglow) and SA (stationary afterglow) experiments carried out at 50--2000 Pa. 

In our recent study \cite{glosik2008rhi} we have shown that at 260~K a significant fraction of the H$_3^+$ + e$^-$
collisions leads to formation of long-lived (up to tens of picoseconds) rotationally-excited neutral
Rydberg molecules $\mathrm{H}_3^*$. The formation of long-lived $\mathrm{H}_3^*$ and 
collisions of $\mathrm{H}_3^*$ with helium atoms can influence the overall process of
recombination of the $\mathrm{H}_3^+$ dominated afterglow plasma. By measuring
the helium pressure dependence of the $\mathrm{H}_3^+$ recombination rate coefficient we have found that at 260~K the ternary recombination rate is comparable with the binary rate, 
already at pressures of several hundred Pa. The observed ternary recombination of $\mathrm{H}_3^+$ ions is more efficient 
by a factor of 100 than the ternary recombination rate predicted by
Bates and Khare \cite{bates1965rpi}. This suggests that ternary recombination of $\mathrm{H}_3^+$ ions
is associated with intrinsic features of the interaction between $\mathrm{H}_3^+$ and
electrons at low collision energies \cite{glosik2008rhi}. The recombination process described
by Bates  \cite{bates1965rpi} is a ternary process of a different nature.

This study presents a further extension of our measurements to a broader range of temperatures and it extracts the
temperature dependence of the binary and ternary recombination rate coefficients of
$\mathrm{H}_3^+$ ions. Before presenting these new results and their interpretation, some of us (the Prague group) wish to provide a few comments concerning the plasma experiments made in Prague in recent years.

Previously, we have studied the $\mathrm{H}_3^+$ recombination in decaying plasmas formed from a He/Ar/$\mathrm{H}_2$ gas mixture using several afterglow experiments based on several modifications of flowing and stationary afterglow apparatuses (FALP
\cite{glosik2003rha}, AISA--Advanced Integrated Stationary Afterglow \cite{plasil2002ais,glosik2001esr,poterya2002rd}, 
TDT-CRDS--Test Discharge Tube with Cavity Ring-Down Spectroscopy \cite{macko2004ash,macko2004dhd}).  We have
systematically investigated the dependence of the recombination process on hydrogen
partial pressure. Initially (in 2000), our intention was to explain the
influence of the internal excitation of $\mathrm{H}_3^+$ ions formed in a sequence of ion-molecule
reactions on the measured recombination rate coefficient. Experimental conditions were such that the plasma was decaying for a long time (up to 60 ms), so the internal
excitation was quenched in collisions with He and $\mathrm{H}_2$. As a rule,
we skip the first 10 ms of the decay process considering it as a formation time. At low $\mathrm{H}_2$
densities, $[\mathrm{H}_2] < 10^{12}\ \mathrm{cm}^{-3}$, and helium pressures of 100--300~Pa, we have observed
a decrease of the recombination rate coefficient from $\sim 10^{-7}$cm$^3$s$^{-1}$ to $10^{-8}$cm$^3$s$^{-1}$ when the $\mathrm{H}_2$ density was decreased
from  $\sim10^{12}\ \mathrm{cm}^{-3}$ to $\sim 10^{11}\ \mathrm{cm}^{-3}$. At high hydrogen densities, $[\mathrm{H}_2] > 10^{12}\ \mathrm{cm}^{-3}$, the measured recombination rate coefficient was independent of $[\mathrm{H}_2]$. We will consider this again below when discuss equilibrium conditions for recombining H$_3^+$ plasma. A similar dependence on $\mathrm{D}_2$ density was observed in a D$_3^+$-dominated afterglow plasma (see e.g. \cite{plasil2002ais,poterya2002rd}).

In our early experimental publications \cite{glosik2000rhi,Kudrna2000ais}, we derived the recombination rate coefficient from experimental measurements assuming that the process is strictly binary (H$_3^+$ + e$^-$) at the low-pressure limit. The assumption was based on the then-existing level of knowledge about H$_3^+$ recombination (see the recent and older reviews in Refs. \cite{larsson2008drh,johnsen2005,larsson2008,amano1990drr,johnsen1998,adams1984mdr}). However, we soon realized that the
observed dependence on hydrogen density is coupled to the multistep character
of the recombination process in a plasma.
The theory of binary dissociative recombination of H$_3^+$ is more immediately applicable to the storage ring experiments. Therefore we denoted the plasma recombination rate constant as an effective deionization rate constant $\alpha_\mathrm{eff}$ \cite{plasil2002ais,glosik2001esr}. Later experiments using absorption spectroscopy for ion density measurements and for the identification of the recombining ions (TDT-CRDS) \cite{macko2004ash,macko2004dhd,Macko2002hrc} did not clarify the mechanism of recombination in an afterglow plasma.

The interpretation of data from storage ring \cite{larsson2008,kreckel2005hrd,mccall2003mn,strasser2002bda} experiments and from afterglow \cite{plasil2002ais,larsson2008,amano1990drr,Gougousi1995131,smith1993drh,poterya2002rd,glosik2003rha,macko2004ash,adams2006emi} experiments was at this point not reconciled \cite{johnsen2005,larsson2008,larsson2008drh,Glosik:06:Action}. We stress here one principal difference between the two types of experiments: In a storage ring experiment, the recombining ions and electrons have a small adjustable relative velocity; the measured cross-section corresponds to the binary electron-ion recombination process. In a plasma experiment an ambient gas must be used. Typically, the pressure of the ambient gas in the afterglow experiments is 50--2000 Pa. The ions and electrons in afterglow plasma undergo multiple collisions with neutral particles (He and $\mathrm{H}_2$ in the experiments discussed here) prior to their mutual collisions; these are collisions at low energy (around $\sim 0.025\ \mathrm{eV}$). In a storage ring, collisions of H$_3^+$ with neutral particles are also possible, but when an ion collides with a background gas molecule or atom, it is removed from the beam and does not contribute any further to the observable recombination events \cite{larsson2008}. 

Plasma experiments measure the thermally averaged rate constant. It means that the afterglow plasma should ideally be in thermal equilibrium with respect to all degrees  of freedom, in particular with respect to the rotational, vibrational and nuclear spin states of H$_3^+$. Molecular hydrogen present in the recombining plasma equilibrates the ortho/para-H$_3^+$ ratio and should make the ``kinetic'' temperature of ions and electrons to coincide with the He temperature \cite{Korolov:08:Measurements,trunec2003iee}. An internal excitation of $\mathrm{H}_3^+$ ions obtained in the process of H$_3^+$ formation \cite{plasil2002ais,johnsen2005} is also quenched in collisions with He atoms, but the population of the lowest energy levels (lowest ortho and para-states) is governed by collisions with H$_2$. One has to have in mind that proton transfer reactions (from $\mathrm{ArH}^+$ or $\mathrm{H}_2^+$ to $\mathrm{H}_2$) producing $\mathrm{H}_3^+$ occur at nearly the rate of elastic collisions, but the rate coefficients for reactions between $\mathrm{H}_3^+$ and $\mathrm{H}_2$, which change ortho-$\mathrm{H}_3^+$ to para-$\mathrm{H}_3^+$ and vice versa (``state changing collisions'') are a factor of 5--10 smaller, $k_\mathrm{sc}\sim 3\times 10^{-10}\ \mathrm{cm}^3\mathrm{s}^{-1}$ (see Refs.
\cite{cordonnier2000srn,gerlich2006dca,pagani2008cml}). If recombination of $\mathrm{H}_3^+$ is sensitive to the nuclear spin state of $\mathrm{H}_3^+$ then recombination drives the $\mathrm{H}_3^+$ ions out of ortho/para thermal equilibrium. In contrast, the ``state-changing collisions'' with $\mathrm{H}_2$ shift the ortho/para  $\mathrm{H}_3^+$ distribution towards the thermal equilibrium. In this case, thermal equilibrium means a population of rotational states corresponding to thermodynamic equilibrium at the given temperature. Note that we have already demonstrated that in a plasma at 260~K the recombination of ortho and para-$\mathrm{H}_3^+$ ions is very different \cite{glosik2008rhi}. The recombination frequency (number of events per unit time) is
$\nu_\mathrm{rec}\sim\alpha_\mathrm{eff}n_\mathrm{e}$ and the ``state changing frequency'' is $\nu_\mathrm{sc}\sim k_\mathrm{sc} [\mathrm{H}_2]$. The condition $\nu_\mathrm{sc} > \nu_\mathrm{rec}$ required for thermal equilibrium leads to $k_\mathrm{sc} [\mathrm{H}_2] > \alpha_\mathrm{eff} n_{\rm e}$. If values typical for FALP experiments ($n_\mathrm{e}\sim 2\times 10^9\ \mathrm{cm}^{-3}$ and
$\alpha_\mathrm{eff}\sim1.5\times 10^{-7}\ \mathrm{cm}^3\mathrm{s}^{-1}$) are used in this inequality, we obtain the condition on the
hydrogen density: $[\mathrm{H}_2] > 10^{12}\ \mathrm{cm}^{-3}$. This is an important density value for the interpretation of data in the plasma
experiments (see also the definition of the rate coefficient in plasma as it is discussed in \cite{atkins1988}). It is presumably not accidental that the density $10^{12}\ $cm$^{-3}$ is the same as the density at which the measured effective $\mathrm{H}_3^+$ recombination rate coefficient changes its character. We have in mind that at temperatures below 300 K $k_\mathrm{sc}$ and $\alpha_\mathrm{eff}$ are dependent on rotational excitation (ortho and para states) of $\mathrm{H}_2$ and $\mathrm{H}_3^+$ than also the condition for thermal equilibrium is more complex. We will come to this point again later.  

Because a recombining plasma contains neutral molecules and atoms, collisions between ions and electrons are perturbed by the neutral particles, which can influence the observed effective plasma recombination rate. The perturbation could be significant if in a electron-ion collision a long-lived highly excited intermediate neutral molecule is formed (see the discussion in Refs. \cite{glosik2008rhi,johnsen1998}). If the
third particle is an atom from the buffer gas, the probability of such a collision is proportional to its pressure and, as a consequence, the apparent recombination rate coefficient will be pressure dependent. Similar phenomena are well known from studies of ion-molecule association reactions, which can have a binary (e.g. radiative association) and/or  three-body character depending on the lifetime of the  complex. The rate of these processes can depend on pressure and temperature (e.g. associative reactions, see the discussion in \cite{gerlich1992eir}). Third-body perturbation of H$_3^+$ binary recombination in a plasma was already mentioned in our recent study \cite{glosik2008rhi}. In the present study, we show further experimental evidence for the phenomenon and further aspects of the theoretical interpretation, in conjunction with  numerical calculation.

The rest of the article is organized in as follows. First, we briefly describe the experiments in section \ref{sec:exps}. Then we present new 
experimental results in section \ref{sec:exp_res}, where we interpret the observed dependencies of the apparent binary recombination rate coefficients ($\alpha_\mathrm{eff}$) on hydrogen and helium pressure, and from that analysis, derive the binary and ternary recombination rate coefficients. Finally, in section \ref{sec:theory} we introduce our theoretical description of ternary recombination of an H$_3^+$ plasma, present our calculation of the lifetimes of the long-lived neutral molecules $\mathrm{H}_3^*$ formed in the electron-ion collisions, and estimate the rate coefficient for the ternary channel of H$_3^+$ recombination. Section \ref{sec:concl} summarizes our conclusions.

\section{Experiments}
\label{sec:exps}
\begin{figure}
\includegraphics[width=8.6cm]{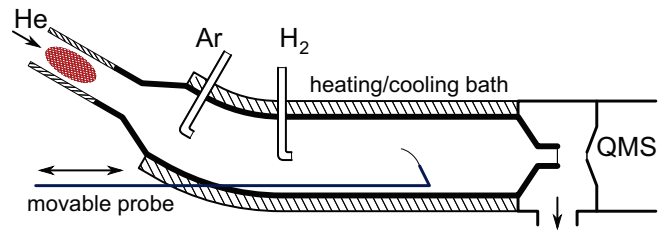}
\caption{Cryo-FALP. Plasma created in a microwave discharge is carried
along the flow tube by helium carrier/buffer gas (from left to right in the
drawing). After addition of Ar (via the indicated gas inlet) the plasma is
converted from $\mathrm{He}_2^+$ dominated to $\mathrm{Ar}^+$ dominated. Further downstream $\mathrm{H}_2$ is
injected to an already-relaxed plasma (with thermalized electrons) and an $\mathrm{H}_3^+$
dominated plasma is formed. The electron density decay downstream from the
hydrogen entry port is monitored by an axially movable Langmuir probe.
\label{fig1}}
\end{figure}

For measurements of pressure and temperature dependences of the $\mathrm{H}_3^+$ 
recombination rate we have built the Cryo-FALP apparatus,
designed to operate in the range 77--300 K and at helium pressures adjustable
from $\sim400$ to $\sim2000$ Pa. UHV technology and high buffer gas purity 
(the level of impurities is $<0.1$ ppm) are used in the Cryo-FALP (Fig.~\ref{fig1}).

\begin{figure}
\includegraphics[width=8.6cm]{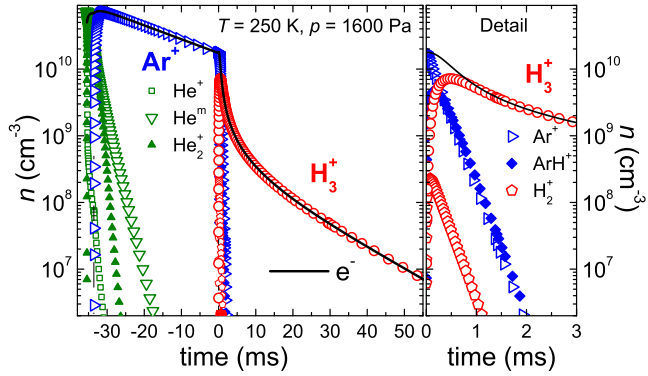}
\caption{Left panel: Calculated dependences of ion formation and plasma decay in FALP after addition of Ar ($7.7\times10^{12}\ \mathrm{cm}^{-3}$) and H$_2$ ($5\times10^{12}\ \mathrm{cm}^{-3}$). 
Argon was added in the flow tube 33~ms before H$_2$. The time scale origin is set at the position of the hydrogen injection port.
The right panel focuses on a narrow time interval corresponding to a transition from $\mathrm{Ar}^+$ dominated to $\mathrm{H}_3^+$ dominated plasma after adding hydrogen. 
\label{fig2}}
\end{figure}

The Cryo-FALP apparatus is a low temperature high pressure variant of the standard FALP apparatus \cite{florescumitchell2006dr}. Here we will just describe the essential features of the new construction. In Cryo-FALP, plasma is created in a microwave discharge (10--30 W) in the upstream glass section of the flow tube (at 300 K). Because of the high pressure, $\mathrm{He}^+$ ions formed by electron impact then react in a three-body association reaction with two atoms of helium, and a $\mathrm{He}_2^+$ dominated plasma is formed.
Downstream from the discharge region, Ar is added to $\mathrm{He}_2^+$ dominated afterglow plasma to remove helium metastable atoms (He$^m$ in Fig.~\ref{fig2})  formed in the discharge and to form $\mathrm{Ar}^+$ dominated plasma (see details in \cite{plasil2002ais,johnsen2005,glosik2000rhi,glosik2001esr}). Via the second entry port situated approximately 35 ms downstream from the Ar entry port, hydrogen (diluted in He) is introduced into the plasma, which at this point is already cold. Note that the position in the flow tube is linked to the decay time. In a sequence of ion-molecule reactions, an $\mathrm{H}_3^+$ dominated plasma is formed shortly after the second entry port.

We have used a numerical model to simulate the process of formation of $\mathrm{H}_3^+$ dominated plasma. Examples of ion density evolutions calculated within the model for conditions typical for the present Cryo-FALP experiments are shown in Fig.~\ref{fig2}. Similar calculations were carried out for all of our experiments presented in this work. Argon also plays a role in plasma relaxation and in formation of $\mathrm{H}_3^+$ ions. However, when an $\mathrm{H}_3^+$ dominated plasma has already been formed, Ar does not play any role (in contrast with He) because its density is at least four orders of magnitude lower than the helium density. Downstream from the Ar entry port the flow tube is thermally insulated and cooled to the desired temperature by liquid nitrogen. Because theory suggests a very strong temperature dependence of the process of interest, we monitored carefully the temperature of the flow tube. An axially movable Langmuir probe \cite{swift1970} is used to measure the electron density decay downstream from the hydrogen entry port \cite{glosik2003rha}. Electron energy distribution functions (EEDF) were checked along the flow tube to characterize plasma relaxation during the afterglow time \cite{Korolov:08:Measurements,plasil2009nme}. Recombination of $\mathrm{O}_2^+$ (``benchmark ion'' \cite{plasil2002ais,johnsen2005}) was used for calibration of the Langmuir probe over a broad range of pressures. 

An advanced analysis was used to fit the decay curves \cite{plasil2002ais,poterya2002rd,Korolov:08:Application}
with the purpose to obtain recombination rate coefficients from the measured decay of electron densities. In the analysis
we have assumed that once an $\mathrm{H}_3^+$ dominated plasma is formed, the plasma decay can
be described by a single value of the recombination rate coefficient, which we call the
effective apparent binary recombination rate coefficient, $\alpha_\mathrm{eff}$. The plasma
decay can then be described by the balance equation with a recombination and
a diffusion loss terms:
\begin{equation}
\frac{\de n_\mathrm{e}}{\de t} = -\alpha_\mathrm{eff}n_\mathrm{e}n_+ - 
\frac{n_\mathrm{e}}{\tau_\mathrm{D}} = -\alpha_\mrm{eff}n_\mrm{e}^2-
\frac{n_\mrm{e}}{\tau_\mrm{D}}\,,
\label{eq1}
\end{equation}
where $n_\mathrm{e}$ and $n_+$ are electron and ion densities, respectively. We assume
that plasma is quasineutral ($n_\mathrm{e} = n_+$). The constant $\tau_\mathrm{D}$ characterizes the
ambipolar diffusion during the afterglow. 
The recombination of $\mathrm{H}_3^+$ ions at temperatures below 300~K depends strongly on the rotational excitation of ions.
The assumption about the constant value of $\alpha_\mathrm{eff}$ must be discussed for particular experimental conditions.

In the present experiments we use helium densities in the range $[\mathrm{He}]\sim 6\times 10^{16}\text{ -- }6\times 10^{17}\ \mathrm{cm}^{-3}$ and hydrogen densities in the range $[\mathrm{H}_2]\sim10^{11}\text{ -- }10^{14}\ \mathrm{cm}^{-3}$. In the Cryo-FALP and other experiments discussed we use normal hydrogen (n-$\mathrm{H}_2$) i.e. the mixture of ortho and para-$\mathrm{H}_2$ corresponding to 300~K (approximately 25\% of para $\mathrm{H}_2$). The variation of the ortho/para-H$_2$ ratio  with temperature in the interval 77~K--300~K is not significant for the results of the present experiments. (In thermal equilibrium at 100~K the fraction of para~$\mathrm{H}_2$ is $\sim38\%$, and at 77~K the fraction is $\sim50\%$).

\section{Experimental results}
\label{sec:exp_res}

\begin{figure}
\includegraphics[width=8.6cm]{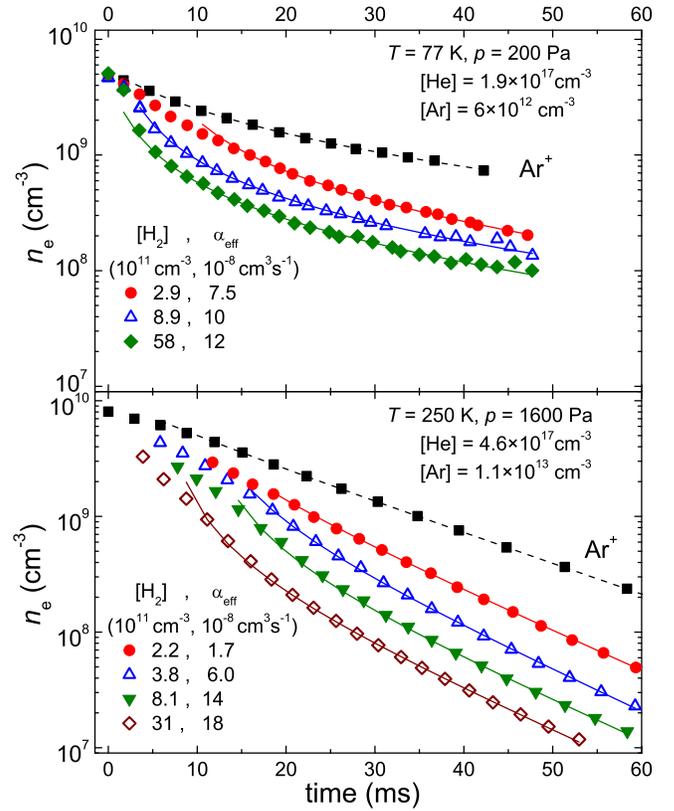}
\caption{Examples of electron density decay curves measured in the $\mathrm{H}_3^+$ dominated
plasma at several hydrogen densities. Upper panel: Cryo-FALP experiment.
Lower panel:  FALP experiment. The effective recombination rate coefficients obtained at
different hydrogen densities are indicated. For comparison, both panels also show the decay curves (rectangles) obtained in the $\mathrm{Ar}^+$ dominated plasma with $[\mathrm{H}_2]=0$.
\label{fig3}}
\end{figure}

We monitored decay of the afterglow plasma in a He/Ar/$\mathrm{H}_2$ mixture at different temperatures and over a broad range of helium and hydrogen densities. Examples of decay curves measured at 77~K and at 250~K at several hydrogen densities are plotted in Fig.~\ref{fig3}. The dependence of the decay rate on hydrogen density is evident. The obtained apparent recombination rate coefficients ($\alpha_\mathrm{eff}$) depend on all three parameters $\alpha_\mathrm{eff}$($T$, [H$_2$], [He]). It clearly indicates that the observed ``deionization process'' is not pure binary dissociative recombination.  In Fig.~\ref{fig3} we have also plotted the decay curves measured in a He/Ar afterglow plasma dominated by $\mathrm{Ar}^+$ ions at 77~K and at 250~K in otherwise very similar conditions. At 250~K (see lower panel) the decay curve is exponential (straight line in the semilogarithmic plot) because recombination of these atomic ions is very slow and the decay is governed by ambipolar diffusion. At 77~K (see upper panel of Fig.~\ref{fig3}) we observe a faster decay during the early afterglow (at higher electron densities). We assume that this faster decay is primarily due to collisional radiative recombination (CRR) \cite{mcdaniel1993,skrzypkowski2004fam} and partly also due to the formation of $\mathrm{Ar}_2^+$ ions (in three-body association at low temperatures) followed by immediate recombination of these ions \cite{Pysanenko:02:Recombination}. The rate of the decay is comparable with the rate calculated for the CRR at $\sim85$~K. 

\begin{figure}
\includegraphics[width=8.6cm]{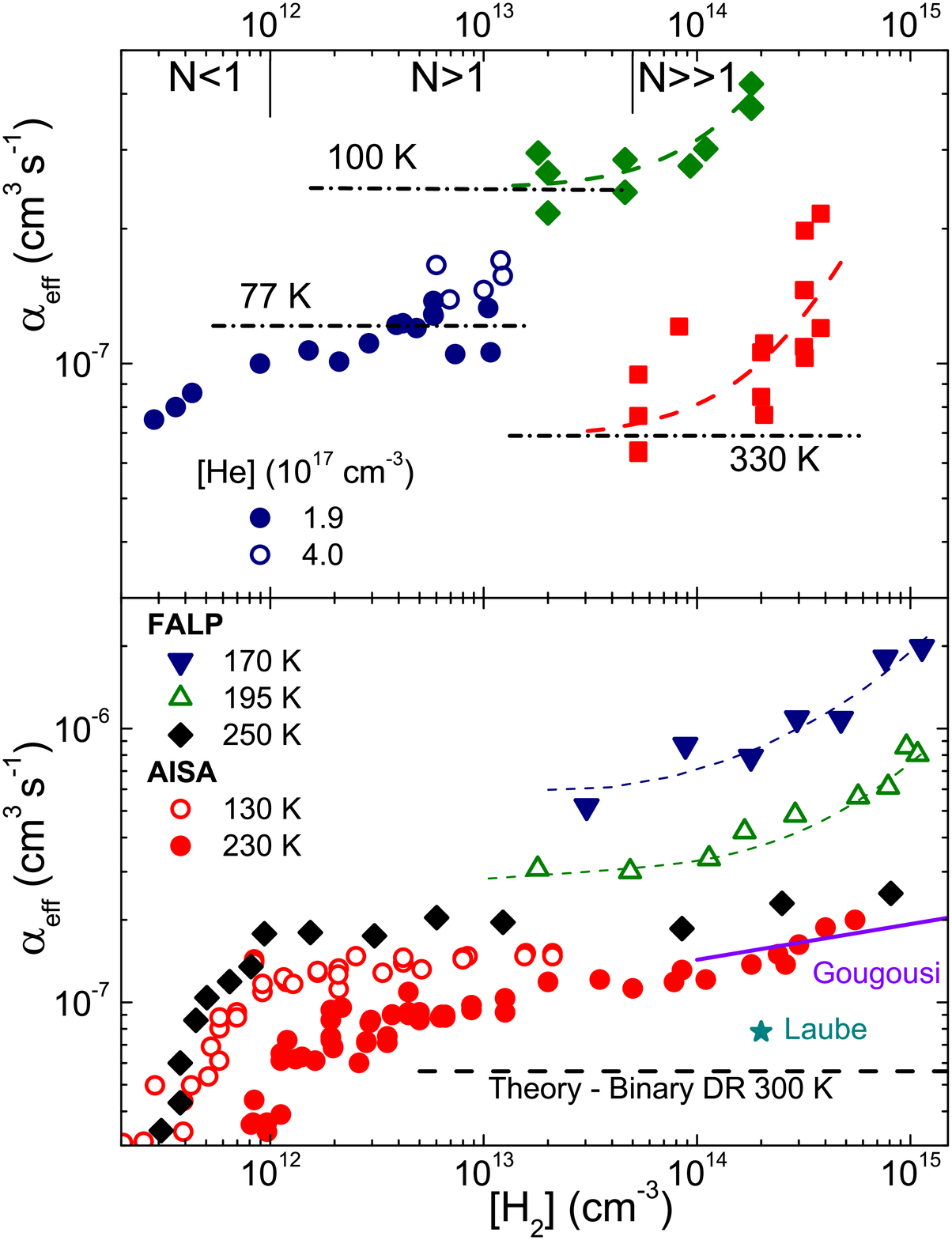}
\caption{Measured dependencies of the effective recombination rate
coefficient of $\mathrm{H}_3^+$ ions on hydrogen density. The data obtained by Cryo-FALP and
TDT-CRDS are plotted in the upper panel. The data obtained in other experiments are
plotted in the lower panel. We have also plotted examples of data obtained in our
previous studies using: AISA, TDT-CRDS and FALP. The FALP data by Laube \textit{et al.}
\cite{laube1998nfm} and Gougousi \textit{et al.} \cite{Gougousi1995131} are plotted for comparison. Additionally, we show the
value of the theoretical recombination rate coefficient calculated for binary dissociative recombination (DR), $\alpha_\mathrm{Bin}(300\ \mathrm{K})$ \cite{dossantos2007drh,pagani2008cml}. Note that for this theoretical result, only the binary DR process is considered, and as such it cannot depend on the hydrogen
density.
\label{fig4}}
\end{figure}

The apparent binary recombination rate coefficients ($\alpha_\mathrm{eff}$) obtained from measured decay curves in the H$_3^+$ dominated plasma are plotted as functions of hydrogen density in Fig.~\ref{fig4}. Examples of the data obtained in other experiments (FALP, AISA and TDT-CRDS) are also included in Fig.~\ref{fig4}. Below we summarize the data plotted in Fig.~\ref{fig4}.

Upper panel:
\begin{enumerate}
\item 77 K, Cryo-FALP. At a fixed flow tube temperature ($T$ = 77~K) and at fixed $[\mathrm{He}]=1.9\times 10^{17}\ \mathrm{cm}^{-3}$, the dependence of $\alpha_\mathrm{eff}$ on $\mathrm{H}_2$ density was measured from $[\mathrm{H}_2]\sim2\times 10^{11}\ \mathrm{cm}^{-3}$ to $\sim10^{13}\ \mathrm{cm}^{-3}$. Then, for several other helium densities the recombination rate coefficient was only measured in a limited range of hydrogen densities close to $[\mathrm{H}_2]\sim10^{13}\ \mathrm{cm}^{-3}$. We plot only a few examples. 
\item 100 K, TDT-CRDS. The $\mathrm{H}_3^+$ ion density was measured by using laser absorption spectroscopy (CRDS technique). During the discharge pulse and during the recombination dominated afterglow, the ion temperature was determined from the Doppler broadening of an absorption line. The details of such experiments can be found in \cite{glosik2005rsh}. 
\item 330 K, TDT-CRDS. The absorption spectroscopy technique (CRDS) was used to measure the temperature and density of the recombining $\mathrm{H}_3^+$ ions \cite{macko2004ash,macko2004dhd,glosik2005rsh}. Relatively high hydrogen density was used in these experiments. Therefore, an extrapolation had to be carried out in order to obtain $\alpha_\mathrm{eff}$ for lower hydrogen densities (see the discussion below). For details of this extrapolation see the discussion in \cite{glosik2003rha,Pysanenko:02:Recombination}. In some experiments, the He temperature was determined by measuring the Doppler broadening of the $\mathrm{H_2O}$ line \cite{glosik2005rsh}. We have also assumed that the temperatures of $\mathrm{H_2O}$ and He are identical. The $\mathrm{H_2O}$ density is very low but sensitivity of CRDS is very high for this molecule.
\end{enumerate}

Lower panel: 
\begin{enumerate}
\setcounter{enumi}{3}
\item 130 and 230~K, AISA. Examples of data measured with AISA. The data were measured at a fixed temperature and at a fixed pressure as a function of hydrogen density.
\item 170, 195, 250~K, FALP. Three different versions of FALP designed to work at high He pressures were used in these experiments.
\item 300~K, Experiment by Laube \textit{et al.} \cite{laube1998nfm}. Low pressure FALP experiment: [$\mathrm{H}_2$] = 2$\times 10^{14}\ $cm$^{-3}$, [He] = $1.6\times 10^{16}\ $cm$^{-3}$, obtained $\alpha_\mathrm{eff}$ is $7.8\times10^{-8}\ $cm$^{3}$s$^{-1}$.
\item 295~K, experiment by Gougousi \textit{et al.} \cite{Gougousi1995131}. In this FALP experiment the dependence of the recombination rate coefficient on [$\mathrm{H}_2$] was observed. The helium pressure was about 130 Pa.
\item 300~K, theory \cite{dossantos2007drh,pagani2008cml}. The value calculated for binary dissociative recombination (for [H$_2$] = 0).
\end{enumerate}

Notice that in certain experiments the rate coefficients were measured over a limited range of hydrogen densities. In our experiments, we have covered the $10^{10}\ \mathrm{cm}^{-3} < [\mathrm{H}_2] < 10^{16}\ \mathrm{cm}^{-3}$ range (AISA, FALP, Cryo-FALP, TDT-CRDS).

In the measured dependences shown in Fig.~\ref{fig4} we distinguish three clearly different
regions of hydrogen densities that exhibit specific behavior of $\alpha_\mathrm{eff}$ as a function of [$\mathrm{H}_2$]. We
indicate these regions as $N<1$, $N>1$, and  $N\gg1$. The ``vertical shifts''  for some dependences in
Fig.~\ref{fig4} (such as the data presented with open and full triangles in the lower panel) are discussed below. We characterize the three regions as follows.
\begin{enumerate}
\item $[\mathrm{H}_2] < 10^{12}\ \mathrm{cm}^{-3}$, $N<1$. At these hydrogen densities $\alpha_\mathrm{eff}$ decreases with decreasing hydrogen density. At such conditions the $\mathrm{H}_3^+$ ions formed by an exothermic proton transfer (from $\mathrm{ArH}^+$ or $\mathrm{H}_2^+$ to $\mathrm{H}_2$) do not have enough collisions with $\mathrm{H}_2$ to establish ortho/para-H$_3^+$ thermal equilibrium prior to their recombination (see e.g. the discussion in \cite{cordonnier2000srn,gerlich2006dca,pagani2008cml}). The number of these H$_2$ + H$_3^+$ collisions that a $\mathrm{H}_3^+$ ion will undergo prior to its recombination with an electron (at typical conditions of the discussed afterglow experiments) is indicated in the Fig.~\ref{fig4} by number $N$. If $N < 1$, the decay of the plasma caused by the recombination (at a given electron density) is faster than the rate of rethermalization. Therefore, in this non-equilibrium regime, the individual state composition of H$_3^+$ afterglow plasma is different along the flow tube because the effective recombination rate depends on the absolute value of electron density, which varies along the length of the tube. By ``state composition'' we mean not only the kinetic energy distribution, which is established in collisions with He atoms with a nearly collisional rate, but also the rotational and ortho-H$_3^+$/para-H$_3^+$ state distribution. A quantitative description of this particular regime would require a much deeper theoretical analysis.
\item $10^{12}\ \mathrm{cm}^{-3} < [\mathrm{H}_2]<5\times 10^{13}\ \mathrm{cm}^{-3}$, $N>1$. In this regime, the measured rate coefficients are nearly independent of [$\mathrm{H}_2$]. On the basis of the same arguments mentioned above, it is clear that at $[\mathrm{H}_2]>10^{12}\ \mathrm{cm}^{-3}$, the $\mathrm{H}_3^+$ ion (formed shortly after hydrogen is injected) has several collisions with $\mathrm{H}_2$ prior to its recombination with an electron. Because only some collisions lead to a change in rotational excitation of $\mathrm{H}_3^+$ or  in ortho$\leftrightarrow$para transitions \cite{cordonnier2000srn,gerlich2006dca}), we assume that if $[\mathrm{H}_2]>10^{12}\ \mathrm{cm}^{-3}$ the recombining ions in the flow tube will be in thermal equilibrium corresponding to the hydrogen temperature, which is assumed to be equal to the helium temperature. Because of the independence of $\alpha_\mathrm{eff}$ on [$\mathrm{H}_2$] we will call this region the ``saturated region''. The boundaries of the saturated region depend on  actual helium pressure, temperature and electron density. The region is broad enough to find conditions where the value of $\alpha_\mathrm{eff}$ is nearly constant (the plateau part of the $\alpha_\mathrm{eff}$([H$_2$]) dependence). Depending on experimental conditions we have covered either the whole saturated region or in some cases just a part of it.
\item $[\mathrm{H}_2]>5\times 10^{13}\ \mathrm{cm}^{-3}$, $N\gg1$. Here the measured recombination rate coefficient increases with hydrogen pressure. This behavior is caused by a formation of H$_5^+$ ions and their fast recombination. The process is temperature and pressure dependent (see details in \cite{glosik2003rha,Novotny:06:The,leu1973mre,hiraoka1975dsh}). Because of the strong temperature and pressure dependence of ternary association, the onset of this region depends on these parameters.
\end{enumerate}

At first sight the ``vertical shifts'' of the dependences plotted in Fig.~\ref{fig4}
are very chaotic. We demonstrate below that they arise because the apparent binary
recombination rate coefficient ($\alpha_\mathrm{eff}$) depends not only on  hydrogen density
and temperature but also on the helium density. In addition, we also show that the temperature dependence of $\alpha_\mathrm{eff}$ is not monotonic.

We will not discuss the $N<1$ region here. However, in connection with the data obtained at 77~K we want to point out one difference from the data obtained at higher temperatures. In all high temperature experiments made at $[\mathrm{H}_2]<10^{12}\ \mathrm{cm}^{-3}$, we have observed a fast decrease of $\alpha_\mathrm{eff}$ with decreasing [$\mathrm{H}_2$] (see e.g. the FALP data measured at 250~K in the lower panel of Fig.~\ref{fig4}). In measurements at 77~K using Cryo-FALP (see the upper panel of Fig.~\ref{fig4}) the decrease is substantially smaller. 
The difference can be partly associated with the effect of collisional radiative recombination (CRR) at 77~K. 
At $[\mathrm{H}_2] \sim10^{13}\ \mathrm{cm}^{-3}$ when the overall recombination rate coefficient  $\alpha_\mathrm{eff}(77\ \mathrm{K})>1.0\times 10^{-7}\ \mathrm{cm^3s^{-1}}$ we can neglect the influence of these processes (see the upper panels of Fig.~\ref{fig3} and Fig.~\ref{fig4}).

\begin{figure}
\includegraphics[width=8.6cm]{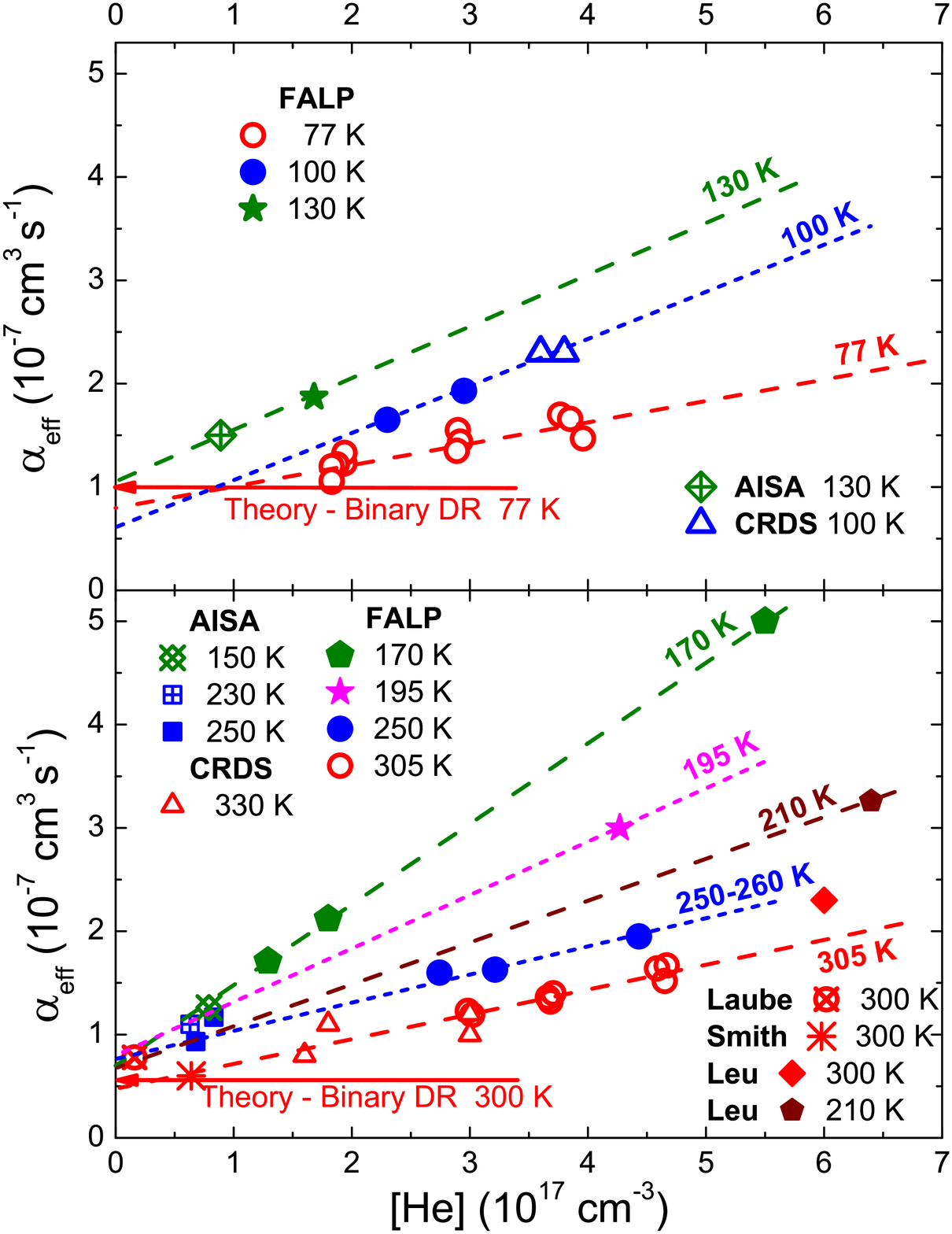}
\caption{The effective binary recombination rate coefficients ($\alpha_\mathrm{eff}$) measured
at the stated temperatures are shown as functions of the helium density. 
 Upper panel: Low
temperature data -- Cryo-FALP (77 K), TDT-CRDS \& Cryo-FALP (100 K) and AISA
\& FALP (130 K). The horizontal line indicates the theoretical value for binary
dissociative recombination at 77 K \cite{dossantos2007drh,pagani2008cml}.
For details, see the summary in the text.
Lower panel: Higher temperature data -- Cryo-FALP (305 K), FALP (250--260
K, 170 and 195 K), TDT-CRDS (330 K), AISA (250--260, 230 and 150 K). Individual
points measured in other laboratories: Smith and Spanel \cite{smith1993drh}, Laube \textit{et al.} \cite{laube1998nfm},
Leu \textit{et al.} \cite{leu1973mre} (see description in the text). The horizontal line indicates the
theoretical value for the binary dissociative recombination of $\mathrm{H}_3^+$ at 300~K
\cite{dossantos2007drh,pagani2008cml}.\label{fig5}}
\end{figure}

Figure \ref{fig4} has a great deal of information, because it actually shows $\alpha_\mathrm{eff}$ as a function of three variables $T$, [H$_2$], and [He]. 
By analyzing the data we found the form of the function $\alpha_\mathrm{eff}$($T$,[H$_2$],[He]) in the ``saturated region'', i.e. for the plasma where the $\mathrm{H}_3^+$ ions before recombining undergo a sufficient number of ``state changing collisions'' with $\mathrm{H}_2$ to reach thermal equilibrium. 
For a better analysis of the experimental data we plotted $\alpha_\mathrm{eff}$ measured at a fixed temperature as a function of helium density. The rate coefficients measured in the three principally different afterglow experiments at several different temperatures at hydrogen densities corresponding to the saturated region are plotted as functions of helium density in Fig.~\ref{fig5}.

We briefly summarize the data plotted in Fig.~\ref{fig5}. Upper panel:
\begin{enumerate}
\item 77~K, Cryo-FALP. The data from measurements at $[\mathrm{H}_2]\sim10^{13}\ \mathrm{cm}^{-3}$ at different $[\mathrm{He}]$ (see upper panel of  Fig.~\ref{fig4}). The experiment was intended as a measurement of the pressure dependence at a fixed temperature. The straight line is the best fit through the measured data.
\item 100 K, TDT-CRDS. The values plotted were obtained from the measured dependences of $\alpha_\mathrm{eff}$ on hydrogen density as limits approaching the saturated region (see upper panel of Fig.~\ref{fig4}). More details are given in \cite{plasil2005rsi}.
\item 100~K, Cryo-FALP. The present measurements.
\item 130~K, AISA. The data were obtained from the measured dependence of $\alpha_\mathrm{eff}$ on hydrogen density (see lower panel of Fig.~\ref{fig4}). More details is given in \cite{glosik2005rsh}.
\item 130~K, Cryo-FALP. The present measurements.
\item 77~K Theory. The horizontal line indicates the theoretical value of the recombination rate coefficient for binary dissociative recombination (DR) of $\mathrm{H}_3^+$ at 77~K \cite{dossantos2007drh,pagani2008cml}.
\end{enumerate}

Lower panel of Fig.~\ref{fig5}:
\begin{enumerate}
\setcounter{enumi}{6}
\item 150, 230 and 250~K, AISA. The values shown were obtained from measured dependencies of $\alpha_\mathrm{eff}$ on hydrogen density as limits approaching the saturated region (see upper panel of Fig.~\ref{fig4} and Ref. \cite{glosik2005rsh}). The accuracy of the values is high because the values were obtained from a number of measurements.
\item 170 and 195~K, FALP \& AISA. The FALP data were obtained similarly to the AISA data, i.e. as limits approaching the saturated region. The straight lines connect the measured FALP points with the AISA points, which are obtained by an extrapolation of the data measured by AISA at 130--230~K.
\item 250--260~K, AISA \& FALP. Compilation of data from several experiments (for details see Refs. \cite{glosik2008rhi,Korolov2009144}).
\item 305~K, Cryo-FALP. The data were obtained by directly changing the helium pressure in the flow tube.
\item 330~K, TDT-CRDS. The values were obtained as limits approaching the saturated region (see upper panel of Fig.~\ref{fig4}).
\item 210 and 300~K, Leu \textit{et al.} \cite{leu1973mre}. In the experiment a microwave technique was used to measure the electron density. The values $2.3\times 10^{-7}\ \mathrm{cm^{3}s^{-1}}$ and $3.3\times 10^{-7}\ \mathrm{cm^{3}s^{-1}}$ for 300~K and 210~K respectively (pressure $\sim 2.6$~kPa) were taken from Figs. 2 and 4 of Ref. \cite{leu1973mre}.
\item 300~K, Laube \textit{et al.} \cite{laube1998nfm}, a FALP (FALP-MS) experiment. The measured value is $\alpha_\mathrm{eff} = 7.8\times 10^{-8}\ \mathrm{cm^{3}s^{-1}}$ (see Fig.~\ref{fig4}). He pressure was $\sim 70$~Pa. Used hydrogen density corresponds to the saturated region.
\item 300~K, Smith \& Spanel \cite{smith1993drh} a FALP experiment. We show the value from their plot of the recombination rate coefficient as a function of position along the flow tube (in Fig.~\ref{fig4}). In the conditions that arise shortly after injection of hydrogen, Smith \& Spanel measured the
value $\alpha_\mathrm{eff} \sim6\times 10^{-8}\ \mathrm{cm^{3}s^{-1}}$ for a relatively long time (see Fig.~\ref{fig4} in \cite{smith1993drh}). Further downstream, they obtained a lower value of the recombination rate coefficient. The helium pressure was $\sim 260$~Pa.
\item 300~K, Gougousi \textit{et al.} \cite{Gougousi1995131}. We did not include their value in the graph because their measurements were at hydrogen densities too high and therefore out of the saturated region. We only mention their experiment in order to show that we included this study in our considerations.
\end{enumerate}

With regard to data obtained by Laube \textit{et al.} \cite{laube1998nfm}, Smith \& Spanel \cite{smith1993drh}, and Gougousi \textit{et al.} \cite{Gougousi1995131}, we have not analyzed their experiments in full detail. However, it is clear from their studies that, in agreement with our experimental data, there is a general trend: The effective rate constant $\alpha_\mathrm{eff}$ increases with the increase of helium density (in the temperature interval covered). In Refs. \cite{laube1998nfm,smith1993drh,Gougousi1995131} the authors used relatively low helium pressures. Therefore, the increase of $\alpha_\mathrm{eff}$ with [He] was not as large as we observe in our measurements, such as Cryo-FALP. Very high pressure was used in \cite{leu1973mre}. As a result, they obtained large $\alpha_\mathrm{eff}$ in agreement with the trend.

The experimental data plotted in Fig.~\ref{fig5} show that the apparent effective binary recombination rate coefficient $\alpha_\mathrm{eff}$ measured at a fixed temperature depends linearly on helium density. We will discuss a possible recombination mechanism below, but at
this point we can assume that the process has a binary character at very low
[He], whereas with increasing [He] the helium assisted ternary process begins to contribute substantially to the overall recombination deionization process.
Therefore, we can write for the observed linear dependence
\begin{equation}
\alpha_\mrm{eff} = \alpha_\mrm{Bin}(T) + K_\mrm{He}(T)  [\mrm{He}]
\label{eq2}
\end{equation}
in terms of the rate coefficient $\alpha_\mathrm{Bin}(T)$  for binary recombination and the rate
coefficient $K_\mathrm{He}(T)$ for ternary He-assisted recombination. The two coefficients
depend on temperature. Note that for some temperatures we have the FALP data (e.g. 77~K), for others we used AISA \& FALP data, and also the TDT-CRDS data. The AISA data were obtained at low helium densities. The FALP data were taken at high helium densities and in most cases we could vary the He pressure in the FALP experiments. We extrapolate the data from AISA to temperatures 170~K and 195~K. Then we plot a straight line through these new points (obtained by the extrapolation) and the corresponding points measured by FALP (see lower panel in Fig.~\ref{fig5}). Using the straight line fits we obtain experimental values for $\alpha_\mathrm{Bin}$ and $K_\mathrm{He}$ at 170~K and 195~K.

\begin{figure}
\includegraphics[width=8.6cm]{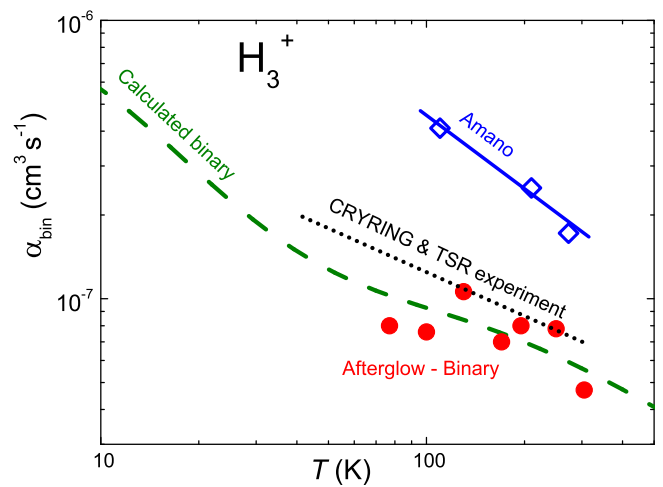}
\caption{The $\mathrm{H}_3^+$  binary recombination rate coefficient measured in the present study (circles). The theoretical values shown (the dashed line) are  calculated for relative populations of para and ortho $\mathrm{H}_3^+$ corresponding to the thermal distribution at the stated $T$ \cite{dossantos2007drh,pagani2008cml}. The dotted line indicates values calculated from cross sections obtained in storage ring experiments \cite{kreckel2005hrd,mccall2003mn,mccall2004drr} using cold ion sources. The present values are obtained as limits approaching the saturated region at low helium density from plots in Fig.~\ref{fig5}. We also show the data (squares) obtained by Amano in pure hydrogen \cite{amano1990drr}. Further details are given in the text. 
\label{fig6}}
\end{figure}

The obtained values of the binary recombination rate coefficients $\alpha_\mathrm{Bin}(T)$ are plotted in Fig.~\ref{fig6} as a function of temperature. We also show on Fig.~\ref{fig6} thermal rate coefficients calculated for binary dissociative recombination of $\mathrm{H}_3^+$ \cite{dossantos2007drh,pagani2008cml}. The agreement is very good. Note that the data for 170~K and 195~K (at high [He]) were published \cite{glosik2003rha,glosik2005rsh} before the calculation \cite{dossantos2007drh}. The data for 250~K was partially obtained \cite{plasil2002ais} also before the calculations. The pressure dependence of the 250~K data set was measured after \cite{glosik2008rhi} the calculations were published. Some data for 100~K and 300~K were also measured earlier, using CRDS \cite{macko2004ash,plasil2005rsi}.

Figure \ref{fig6} also presents the data obtained by Amano using absorption spectroscopy \cite{amano1990drr} in experiments made with pure hydrogen at $\sim 40$~Pa ([H$_2$]$\sim 10^{16}\ $cm$^{-3}$) used as a buffer gas. The large values of the measured recombination rate coefficients indicate that the molecular hydrogen is a more effective three body partner (with an effective ternary rate constant of order of 10$^{-23}\ $cm$^6$s$^{-1}$) than helium, which is not surprising because H$_2$ has rotational and vibrational degrees of freedom.

We have also checked that the observed linear pressure dependence of $\alpha_\mathrm{eff}$  is not associated with the Langmuir probe operating at different pressures from 40 to 2600~Pa. For this test we have used the same probe to measure the recombination rate coefficient for $\mathrm{O}_2^+$ ions \cite{plasil2002ais,johnsen2005}. Our assumption was that $\alpha_{\mathrm{O}_2^+}$ is pressure independent because O$_2^+$ recombination is a direct binary (electron--ion) process. We have also studied the pressure dependence of recombination rate coefficients for $\mathrm{HCO}^+$ and $\mathrm{DCO}^+$ ions \cite{Korolov2009144} using the probe. For both cases (O$_2^+$ and HCO$^+$/DCO$^+$) we observed no pressure dependence. Our $\mathrm{HCO}^+$ and  $\mathrm{DCO}^+$ rate coefficients are in good agreement with results of Leu \textit{et al.} \cite{leu1973mre} and Amano \cite{amano1990drr} measured by different techniques. 
Leu \textit{et al.} \cite{leu1973mre} used a microwave technique to determine electron densities but obtained results are consistent with our observation.

\begin{figure}
\includegraphics[width=8.6cm]{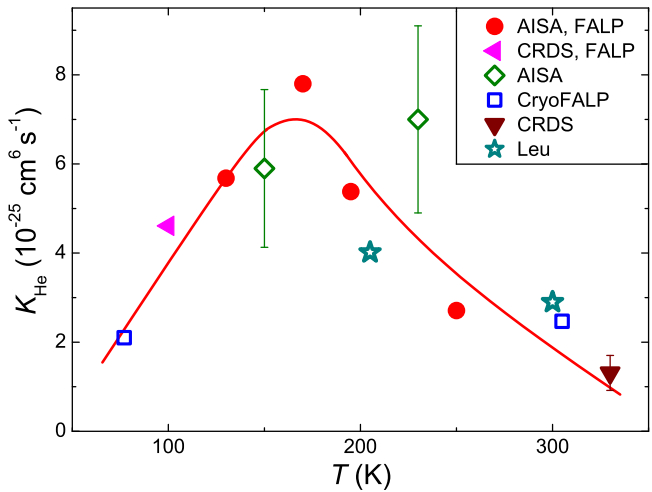}
\caption{The measured ternary recombination rate coefficient, $K_\mathrm{He}(T)$, for helium assisted recombination of $\mathrm{H}_3^+$ with electrons. The legend indicates the experiments used to extract the data. The plotted line is shown merely to guide the eye.
\label{fig7}}
\end{figure}

The ternary rate coefficients obtained from the data plotted in the Fig.~\ref{fig5} are presented in Fig.~\ref{fig7} as a function of buffer gas temperature.  The values obtained for 300~K are: $\alpha_\mathrm{Bin}(300\ \mathrm{K})=(4.7\pm1.5)\times 10^{-8}\ \mathrm{cm^{3}s^{-1}}$, $K_\mathrm{He}(300\ \mathrm{K}) = (2.5\pm1.2)\times 10^{-25}\ \mathrm{cm^{6}s^{-1}}$.  The figure shows that the ternary rate coefficient has a maximum at $\sim170$~K. Towards lower temperatures the rate coefficient is decreasing. This is a surprising result if one takes into account studies of ternary association processes. For such processes the ternary rate coefficients decrease monotonically with temperature (see Refs. \cite{adams1983,GLOSIK:95:STUDIES,Glosik:98:Selected}). 

We briefly summarize the data plotted in Figure \ref{fig7}:
\begin{enumerate}
\item 77~K, Cryo-FALP. The data obtained from the pressure dependence measured in the present experiments at
77~K. See upper panels of Figs \ref{fig3}, \ref{fig4}, and \ref{fig5}. The value of $K_\mathrm{He}$ was obtained from measurements that were repeated many times. 
\item 100~K, Cryo-FALP \& TDT-CRDS. The data obtained by a combination of the values measured in two experiments. The TDT-CRDS values are based on measurements of the ion density evolution during
the afterglow using CRDS absorption spectroscopy. (see the upper panels of Figs. \ref{fig4} and \ref{fig5}). In order to calculate the ion density from the absorption signal, thermodynamic equilibrium was assumed. The assumption should be valid at the hydrogen densities used. The Cryo-FALP values were measured in the present experiments (upper panel of Fig.~\ref{fig5}).
\item 130, 170, and 195~K, AISA \& FALP. Data obtained in two experiments (Fig.~\ref{fig5}).
\item 150 and 230~K, AISA. AISA was used to measure the dependence of $\alpha_\mathrm{eff}$ on hydrogen density over a broad range including the ``saturated region'', (lower panel of Fig.~\ref{fig4}). From these measurements, average values of the recombination rate coefficient in the ``saturated region'' have been obtained for several temperatures (see Fig.~\ref{fig5}). These data are very accurate because the measurements have been done many times at a fixed temperature. As we did not measure the dependence on helium pressure in the AISA experiments, we calculated $K_\mathrm{He}$ using Eq. (\ref{eq2}) with the current theoretical value for $\alpha_\mathrm{Bin}(T)$. The data have been obtained at low pressures, 160--330~Pa. Therefore, the ternary rate coefficients extracted from this data have large error bars.
\item 250--260~K, FALP \& AISA. These data were obtained in several experiments. A detailed description was given in our previous publication \cite{glosik2008rhi} but there is one key difference: In that study, \cite{glosik2008rhi} the ``260~K'' plot also included values obtained at temperatures close to but different from 260~K; the data shown there for $\alpha_\mathrm{eff}$(260~K) were recalculated from the measured values, assuming a $T^{-0.5}$ dependence, which is a small correction. Because of the additional experiments with Cryo-FALP the recalculation procedure was not necessary in the present work.  Now it is clear that $\alpha_\mathrm{eff}$ depends on both temperature and pressure; in the vicinity of 260~K the temperature dependence is steeper than $T^{-0.5}$.
\item 305~K, Cryo-FALP. In this experiment the pressure dependence was measured (see Fig.~\ref{fig5}).
\item 330~K, TDT-CRDS. The measurements were similar to those at 100~K (upper panel of Fig.~\ref{fig4} and lower panel of Fig.~\ref{fig5}). Two different absorption lines were used in these studies. We have obtained the same value of the rate coefficient $\alpha_\mathrm{eff}$ using either absorption line.
\item 210 and 300~K, Data from Leu \textit{et al.} \cite{leu1973mre}. (see Fig.~\ref{fig5}). We have utilized the current theoretical binary recombination rate coefficients for 210 and 300~K and Eq. \ref{eq2} to obtain the corresponding ternary rate coefficients.
\end{enumerate}

\section{Theoretical model for helium-assisted neutralization of the afterglow plasma}
\label{sec:theory}

It is possible to estimate theoretically the rate coefficient, Eq. (\ref{eq2}), of the He-assisted recombination of H$_3^+$ with electrons in the following way. To stress that it is an estimated value, we use symbol $K^{3\mathrm{d}}$ for the theoretical coefficient.

He-assisted recombination is treated as a two step process. In electron-$\mathrm{H}_3^+$ scattering, rotational
autoionization resonances of the neutral molecule $\mathrm{H}_3^*$ play an important role
(see, for example Ref. \cite{kokoouline2003utt,PhysRevA.61.033410}). The star next to $\mathrm{H}_3$ refers to the unstable character of these autoionizing resonances. Such resonances with angular momentum $l=1$ could be very quite broad due to the high probability to capture an electron into the $l=1$ partial wave, but they normally contribute relatively little to the two-body, $\mathrm{H}_3^+ + \mathrm{e}^-$ recombination \cite{kokoouline2003utt}, because almost always, such resonances decay back into a free electron and an $\mathrm{H}_3^+$ ion. As we demonstrate below, lifetimes of some of these resonances can be quite long. If, during their lifetime, the $\mathrm{H}_3^*$ molecule collides with a helium atom, the collision can lead to a change in the electronic state of the outer electron or in the rotational state of the $\mathrm{H}_3^+$ ion and therefore, can make the autoionization process impossible (or, at least, much less probable than the dissociation of $\mathrm{H}_3$). The overall rate coefficient for such He-assisted recombination is given by the formula derived in Ref. \cite{glosik2008rhi}: 
\begin{equation}
K^\mrm{3d} = k_l\Delta t\,\alpha_*\;,
\label{eq3}
\end{equation} where $\alpha_*$ is
the rate coefficient for the formation of $\mathrm{H}_3^*$ ; $k_l$ is the rate coefficient for $l$-changing collisions between He and $\mathrm{H}_3^*$ leading to the eventual dissociation of $\mathrm{H}_3^*$. Finally, $\Delta t$ is the delay time in the $\mathrm{H}_3^++\mathrm{e}^-$ collision in the sense introduced by Smith \cite{PhysRev.118.349}): it is an additional time that the electron spends near the ion in the modified Coulomb potential compared to the collisional time in a pure Coulomb potential. The delay time and the coefficient $\alpha_*$ are substantially different from zero only near resonances. Therefore, the three-body rate coefficient discussed above is expected to vary resonantly as a function of collisional energy. 

We stress here that the rate constants $k_l$, and $\alpha_*$ described above depend on the corresponding scattering energies (and defined as usual as cross-sections multiplied by relative velocities). They are not yet thermally averaged over the Maxwell-Boltzmann distribution. Correspondingly, the ternary rate constant $K^\mrm{3d}$ in Eq. (\ref{eq3}) depends on two energies: the
energy $E$ of the $\mathrm{H}_3^++\mathrm{e}^-$ collision due to the dependence of $\Delta t$ and $\alpha_*$ on $E$;
and the energy $E_\mathrm{He}$ of collision between He and $\mathrm{H}_3^*$ due to the dependence of $k_l$
on $E_\mathrm{He}$. In this approach, however, the coefficient $k_l(E_\mathrm{He})$ is considered to be constant over the energy interval of interest. The rate coefficient $\alpha_*$ in our estimation is evaluated as $v\sigma(E)$, where $v=(2E/m)^{1/2}$ is the relative
velocity, $\sigma(E)$ is the cross-section for the process leading to the delay time,
$m$ is the reduced mass of the $\mathrm{H}_3^++\mathrm{e}^-$ system.

For the following discussion, we assume that at a given energy $E$, there could be several open $\mathrm{H}_3^+ + \mathrm{e}^-$ ionization channels ($i=1,2\ldots n_\mathrm{o}$). Such a situation is possible for the conditions of the present experiment. If there are several open channels, the three-body coefficient $K^{3\mathrm{d}}$ above should be averaged over the incident channels and summed over the final ones. If the incident ionization channel (before collision) is denoted by the index $i$, and the final one (following the collision) is $j$, then the corresponding rate coefficient for the three-body recombination during the $i\to j$ inelastic
collision is given by 
\begin{equation}
k_l\Delta t_{ji}\sqrt{2E/m}\sigma_{ji}(E)\,\text{,}
\label{eq4}
\end{equation}
with $\sigma_{ji}(E)$ being the cross-section for the inelastic collision. 
The delay time $\Delta t_{ji}$ for the process is an element of the
delay-time matrix $\Delta \bm{t}$ as it is introduced and discussed by Smith \cite{PhysRev.118.349}:
\begin{equation}
\Delta t_{ji} = \mathfrak{R}\left[-i\hbar\frac{1}{S_{ji}}\frac{\de S_{ji}}{\de E}\right]\,,
\label{eq5}
\end{equation}
 where $S_{ji}$ is an element of the scattering matrix for the $i\to j$ process.
Notice a  difference in conventions in the present paper and Ref. \cite{PhysRev.118.349}: Here,
the first  index in each matrix denotes the final channel. The second
index denotes the incident channel. In Ref. \cite{PhysRev.118.349}, the convention adopted was the
opposite, i.e., the first index $\sim$ incident channel, second index $\sim$ final one.

The cross-section $\sigma_{ji}(E)$ is given by 
\begin{equation}
\sigma_{ji}(E) = \frac{\hbar^2\pi}{2mE}|S_{ji}|^2\,.
\label{eq6}
\end{equation}
 Using equations (\ref{eq4}, \ref{eq5}, \ref{eq6}) and taking the sum
over the final ionization channels $j$, we obtain the three-body rate coefficient
$K_i^{3\mathrm{d}}$ if the initial state of the $\mathrm{H}_3^++\mathrm{e}^-$ system is $i$:
\begin{eqnarray*}
K_i^{3\mrm{d}} &=& k_l\sum_{j=1}^{n_\mathrm{o}}
\mathfrak{R}\left[-i\hbar\frac{1}{S_{ji}}\frac{\de S_{ji}}{\de E}\right]
\sqrt{2E/m}
\frac{\hbar^2\pi}{2mE}|S_{ji}|^2\\
&=& k_l(2E)^{-1/2}m^{-3/2}\hbar^2\pi\sum_{j=1}^{n_\mathrm{o}}
\mathfrak{R}\left[-i\hbar\frac{1}{S_{ji}}\frac{\de S_{ji}}{\de E}\right]
|S_{ji}|^2
\end{eqnarray*}

The sum in the second line can be simplified as \cite{PhysRev.118.349} 
\begin{eqnarray}
\sum_{j=1}^{n_\mathrm{o}}
\mathfrak{R}\left[-i\hbar\frac{1}{S_{ji}}\frac{\de S_{ji}}{\de E}\right] &=& 
\sum_{j=1}^{n_\mathrm{o}}
\mathfrak{R}\left[-i\hbar S^\dagger_{ji}\frac{\de S_{ji}}{\de E}\right]\nonumber\\
&=&-i\hbar\left({S}^+\frac{\de{S}}{\de E}\right)_{ii} = {Q}_{ii}\;,
\label{eq7}
\end{eqnarray}
where $Q_{ii}$ is a
diagonal element of the lifetime matrix introduced by Smith \cite{PhysRev.118.349}
\begin{equation}
Q = -i\hbar S^\dagger\frac{\de S}{\de E}
\label{eq8}
\end{equation}
and the
product $S^\dagger\frac{\de S}{\de E}$ in the above equation is the regular matrix product. Due to the
aforementioned difference in matrix index conventions, the order of the
product is opposite to the one in Ref. \cite{PhysRev.118.349}. Therefore, for $K_i^{3\mrm{d}}$ we have 
\def\Kitd{K_i^{3\mathrm{d}}}
\begin{equation}
K_i^{3\mrm{d}} = k_l  m^{-3/2}\hbar^2\frac{\pi}{\sqrt{2E}} Q_{ii}\;,
\label{eq9}
\end{equation}
or in atomic units
\begin{equation}
K_i^{3\mrm{d}} = k_l \frac{\pi}{\sqrt{2E}} Q_{ii}\;.
\label{eq10}
\end{equation}

\begin{figure}
\includegraphics[width=8.6cm]{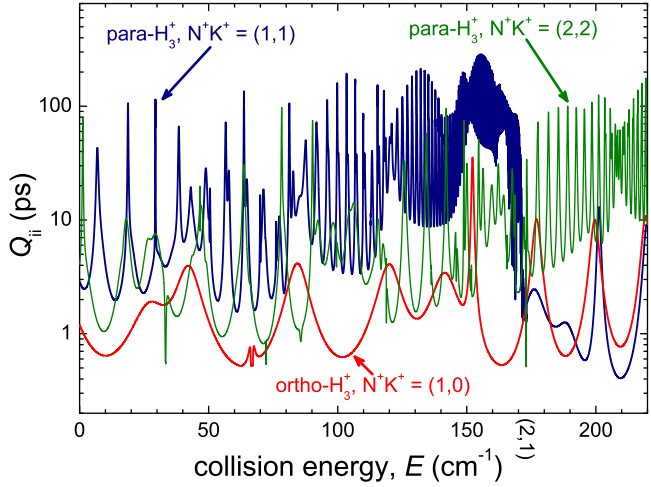}
\caption{Diagonal elements $Q_{ii}$ of matrix $Q$ for the three lowest (rotational)
incident channels for the $\mathrm{e}^-+\mathrm{H}_3^+$ collisions. The rotational channels are
$(N^+,K^+)=(11),(10), \text{ and } (22)$. Each maximum in $Q_{ii}$ corresponds to an
autoionization resonance. The lifetime of a resonances is given by $Q_{ii}/4$
evaluated at the maximum if there is only one channel open, $Q_{ii}=Q$.}
\label{fig8}
\end{figure}

The next step in the evaluation of the three-body rate coefficient is the thermal
average over incident ionization channels $i$ and over the Maxwell-Boltzmann
velocity distribution for a given temperature $T$. The average over incident
channels is given by 
\begin{equation}
\frac{\sum_i\Kitd w_i\exp\left(-\frac{E_i}{k_\mrm{B}T}\right)}{
\sum_i w_i\exp\left(-\frac{E_i}{k_\mrm{B}T}\right)}\;,
\label{eq11}
\end{equation}
where $w_i =(2N_i^++1)(2I_i+1)$ is the statistical weight
of the incident channel, $N_i^+$ and $I_i$ are the angular momentum and the nuclear
spin of the H$_3^+$ ion respectively, $E_i$ is the energy of the incident channel (rotational energy of channel $i$). The average over the velocity (energy)
distribution should be performed over the two energy variables,  $E$ and $E_\mrm{He}$, in
$\Kitd(E, E_\mrm{He})$. Because $\Kitd$ does not depend on $E_\mrm{He}$, the thermal averaging is
reduced to the familiar two-body averaging integral over $E$ only
\begin{equation}
\frac{2}{\sqrt{\pi(k_\mrm{B}T)^{3}}}\int_0^\infty\Kitd(E)\exp\left(-\frac{E}{k_\mrm{B}T}\right)\sqrt{E}\de E\;.
\label{eq12}
\end{equation}
Combining the above equations, we obtain 
\begin{equation}
\left<\Kitd\right> = \frac{
2\sum_i\int_0^\infty\Kitd w_i\exp\left(-\frac{E+E_i}{k_\mrm{B}T}\right)\sqrt{E}\de E}{
\sqrt{\pi(k_\mrm{B}T)^3}\sum_i w_i\exp\left(-\frac{E_i}{k_\mrm{B}T}\right)}\;,
\label{eq13}
\end{equation}
or using the lifetime
matrix element $Q_{ii}$, 
\begin{equation}
\left<\Kitd\right> = \sqrt{\frac{2\pi}{(k_\mrm{B}T)^3}}\frac{
k_l\sum_i\int_0^\infty Q_{ii} w_i\exp\left(-\frac{E+E_i}{k_\mrm{B}T}\right)\de E}{
\sum_i w_i\exp\left(-\frac{E_i}{k_\mrm{B}T}\right)}\;.
\label{eq14}
\end{equation}
The lifetime matrix $Q$ is calculated using Eq. (\ref{eq8})
from the scattering matrices obtained numerically \cite{kokoouline2003utt,dossantos2007drh}. In practice, the
scattering matrix $S^{(N)}$ (and matrix $Q^{(N)}$) are calculated for a fixed value of
the total angular momentum $N$ of the $\mrm{H}_3^++\mrm{e}^-$ system, which is the sum of the
ionic $N^+$ and electronic $l$ angular momenta. The contributions $Q^{(N)}_{ii}$ from
different $N$ should be accounted in Eq. (\ref{eq14}), namely as 
\begin{equation}
Q_{ii} = \frac{1}{2N^++1}\sum_N(2N+1)Q_{ii}^{(N)}\;.
\label{eq15}
\end{equation}
The sum in the
above equation runs over all $N$ for which the incident channel $i$ enters into the
scattering matrix. Since the principal contribution to the rotational capturing
of the electron comes from the $l=1$ partial wave, the sum has three or less
terms.  

\begin{figure}
\includegraphics[width=8.6cm]{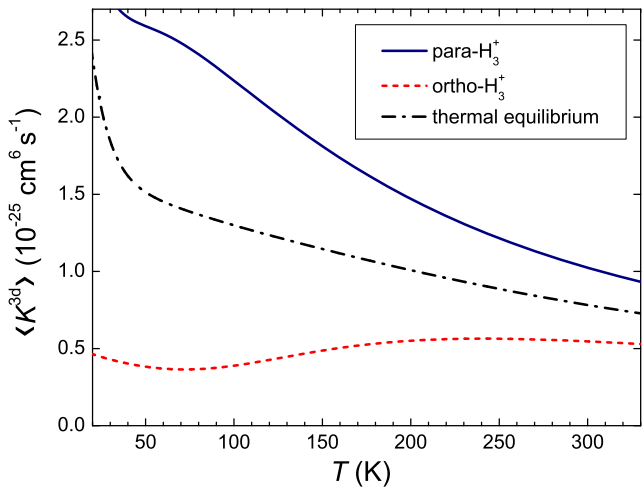}
\caption{Calculated thermally-averaged three-body rate coefficient $\left<K^{3\mathrm{d}}\right>$. The
rate coefficients calculated separately for ortho and para-$\mathrm{H}_3^+$ are very
different. If the recombining plasma is not in thermal equilibrium with respect
to ortho/para ratio, the averaged rate coefficient (dashed) could be very
different from the one shown.\label{fig9}}
\end{figure}

The collisional $l$-changing process is known to be relatively effective
for excited atomic Rydberg states \cite{PhysRevA.18.1339}. Because the $\mathrm{H}_3^*(l=1)$ molecules formed
in the $\mathrm{H}_3^+ +\mathrm{e}^-$ collisions have a large principal quantum number ($n\sim40\text{--}100$, see
Fig.~\ref{fig8}), it is reasonable to use the atomic rates for $l$-changing collisions in
our estimates. For $\mathrm{Na^*+ He}$ $l$-changing collisions \cite{PhysRevA.18.1339} the experimental rate is
$2.3\times 10^{-8}\ \mathrm{cm^{3}s^{-1}}$. In our estimation, we use this value for $k_l$.  Using the
procedure described above and the value above for $k_l$ we have calculated
thermally averaged rate coefficient for the He-assisted recombination of $\mathrm{H}_3^+$.
The coefficient is shown in Fig.~\ref{fig9} as a function of temperature. The overall
agreement with the experimental rate coefficient shown in Fig.~\ref{fig7} is
reasonable given the approximative approach that we used here. Below 150 K, the
experimental curve drops down, but the theoretical (dashed) curve continues to
grow. Interestingly, the curve for pure ortho-$\mathrm{H}_3^+$ (dot-dashed curve) has a
behavior similar to the experiment. In fact, it is possible that in the
experiment the ortho and para-$\mathrm{H}_3^+$ are not in thermal equilibrium at low
temperatures. Our previous calculation \cite{dossantos2007drh} showed that the binary recombination
rate of $\mathrm{H}_3^+$ with electrons is much slower for ortho-$\mathrm{H}_3^+$ than for para-$\mathrm{H}_3^+$: At
low temperatures, due to fast recombination of para-$\mathrm{H}_3^+$ the ratio
ortho/para-$\mathrm{H}_3^+$ could be larger than the one at thermal
ortho/para equilibrium. In such a situation the averaged theoretical rate
coefficient $K^{3\mathrm{d}}$ in Fig.~\ref{fig9} would be closer to the curve for ortho-$\mathrm{H}_3^+$, i.e. it
would be lower at small $T$ similar to the experimental behavior (Fig.~\ref{fig7}).

\section{Conclusions and discussion}
\label{sec:concl}

We have studied recombination of $\mathrm{H}_3^+$ ions in an afterglow plasma, in the presence of a helium buffer gas with a small admixture of molecular hydrogen. The helium densities were in the range $0.5\text{ -- }6\times 10^{17}\ \mathrm{cm}^{-3}$ and hydrogen densities $1\text{--}100\times 10^{12}\ \mathrm{cm}^{-3}$. In such conditions the $\mathrm{H}_3^+$ ions formed in the plasma have several collisions with H$_2$ before they recombine with electrons. Thus, we assume that in these conditions the ions are in ortho/para thermal equilibrium. The apparent binary recombination rate coefficient $\alpha_\mathrm{eff}$ was measured as function of hydrogen and helium densities for several temperatures in the 77--330~K range. From the experimental data we have derived the binary and ternary recombination rate coefficients and their dependences on temperature. The measured binary recombination rate coefficient is in good agreement with recent theoretical calculation over the whole covered temperature range (77--330~K). Therefore, for the first time, the recombination rate coefficients obtained in afterglow plasma experiments agree with storage ring experiments and with theoretical values. As we have demonstrated in the present study, results from previous afterglow plasma experiments were previously interpreted without taking into account the role of the buffer gas and, as a result, those experiments seemed to disagree with the  storage ring experiments and with theoretical calculation. The present work reconcile observation data from the plasma and storage ring experiments and the theoretical result. 

The obtained binary rate coefficient at 300~K is $\alpha_\mathrm{Bin}=(4.7\pm1.5)\times 10^{-8}\ \mathrm{cm^{3}s^{-1}}$. The observed ternary recombination ($K_\mathrm{He}(300\ \mathrm{K}) = (2.5\pm1.2)\times 10^{-25}\ \mathrm{cm^{6}s^{-1}}$) is fast and at pressures of hunderds of Pa it is already dominant over the binary process. The dependence of the measured ternary recombination rate coefficient on temperature has a maximum at $\sim 130$--170~K. The observed ternary process is more effective by factor about $100$ than the process of ternary recombination described by Bates and Khare \cite{bates1965rpi}.

To explain the process of fast ternary recombination we have developed a theoretical model for the process. In particular, we have calculated the delay time in $\mathrm{H_3^++e^-}$ collisions ($\Delta t$ as introduced by Smith \cite{PhysRev.118.349}). We found that the delay time is sensitive to the rotational and nuclear-spin states of the H$_3^+$ ion. The delay time at collision energies $E\sim 150\ \mathrm{cm^{-1}}$ can be of order of 100~ps for para-$\mathrm{H}_3^+$. During that collision time, the $\mathrm{H}_3^*$ molecule ($\mathrm{H_3^++e^-}$ complex) can collide with a helium atom, which enhances the overall plasma neutralization.  The calculated delay time was used to derive the ternary recombination rate coefficient. The derived ternary coefficient as a function of temperature is smaller than the experimental value by a factor of order 2-10, which is plausible agreement for such a sophisticated process (from a theoretical {\it ab initio} point of view) within the somewhat heuristic theoretical method employed. Theory can in principle be further improved. In particular, the coefficient for $l$-changing collisions can be evaluated more accurately. At temperatures below 130~K there is a qualitative difference between measured and calculated values of the ternary rate coefficient: the experimental rate constant decreases with temperature, the theoretical one continues to grow. It is possible that at low temperature ortho and para-H$_3^+$ ion are not in thermal equilibrium in this afterglow plasma because of the very different binary rate constants $\alpha_{\rm Bin}(T)$ that have been predicted for low temperatures. Nevertheless, for a first semiquantitative study of this kind, the agreement between theory and experiment for the ternary rate coefficients is encouraging. 

\begin{acknowledgments}
This work is a part of the research plan MSM 0021620834 financed by the Ministry of Education of the Czech Republic and was partly supported by GACR (202/07/0495, 205/09/1183, 202/09/0642, 202/08/H057) by GAUK 53607, GAUK 124707 and GAUK 86908.  It has also benefitted from support from the National Science Foundation, Grants Nos. PHY-0427460 and PHY-0427376, and from an allocation of NERSC supercomputing resources.
\end{acknowledgments}


\end{document}